\documentclass[aps,prd,nofootinbib,amsmath,amssymb,superscriptaddress,twocolumn,10pt]{revtex4-1}

\pdfoutput=1

\usepackage{graphicx}
\usepackage{dcolumn}
\usepackage{bm}
\usepackage{amssymb}
\usepackage{latexsym}
\usepackage{booktabs}
\usepackage{amsmath}
\usepackage{multirow}
\usepackage{marvosym}
\usepackage{color}
\usepackage[colorlinks=true, linkcolor=red, citecolor=blue]{hyperref}

\newcommand{\be}{\begin{equation}}
\newcommand{\ee}{\end{equation}}
\newcommand{\bq}{\begin{eqnarray}}
\newcommand{\eq}{\end{eqnarray}}

\begin{document}

\title{Prospects for weighing neutrinos in interacting dark energy models using joint observations of gravitational waves and $\gamma$-ray bursts}

\author{Lu Feng}
\affiliation{College of Physical Science and Technology, Shenyang Normal University, Shenyang 110034, China}
\affiliation{Key Laboratory of Cosmology and Astrophysics (Liaoning Province) \& Department of Physics, College of Sciences, Northeastern University, Shenyang 110819, China}
\author{Tao Han}
\affiliation{Key Laboratory of Cosmology and Astrophysics (Liaoning Province) \& Department of Physics, College of Sciences, Northeastern University, Shenyang 110819, China}
\author{Jing-Fei Zhang}
\affiliation{Key Laboratory of Cosmology and Astrophysics (Liaoning Province) \& Department of Physics, College of Sciences, Northeastern University, Shenyang 110819, China}
\author{Xin Zhang}\thanks{Corresponding author.\\zhangxin@mail.neu.edu.cn}
\affiliation{Key Laboratory of Cosmology and Astrophysics (Liaoning Province) \& Department of Physics, College of Sciences, Northeastern University, Shenyang 110819, China}
\affiliation{Key Laboratory of Data Analytics and Optimization for Smart Industry (Ministry of Education), Northeastern University, Shenyang 110819, China}
\affiliation{National Frontiers Science Center for Industrial Intelligence and Systems Optimization, Northeastern University, Shenyang 110819, China}
\begin{abstract}

Cosmological observations can be used to weigh neutrinos, but this method is model-dependent, with results relying on the cosmological model considered. If we consider interactions between dark energy and dark matter, the neutrino mass constraints differ from those derived under the standard model. On the contrary, gravitational wave (GW) standard siren observations can measure absolute cosmological distances, helping to break parameter degeneracies inherent in traditional cosmological observations, thereby improving constraints on neutrino mass. This paper examines the constraints on neutrino mass within interacting dark energy (IDE) models and explores how future GW standard siren observations could enhance these results. For multi-messenger GW observations, we consider the joint observations of binary neutron star mergers by third-generation ground-based GW detectors and short $\gamma$-ray burst observations by missions similar to the THESEUS satellite project. Using current cosmological observations (CMB+BAO+SN), we obtain an upper limit on the neutrino mass in the IDE models of 0.15 (or 0.16) eV. With the inclusion of GW data, the upper limit on the neutrino mass improves to 0.14 eV. This indicates that in the context of IDE models, the improvement in neutrino mass constraints from GW observations is relatively limited. However, GW observations significantly enhance the constraints on other cosmological parameters, such as matter density parameter, the Hubble constant, and coupling strength between dark energy and dark matter.

\end{abstract}
\maketitle

\section{Introduction}
\label{sec1}

Experiments involving solar and atmospheric neutrinos have demonstrated that neutrinos possess mass and exhibit substantial mixing among different species \cite{Lesgourgues:2006nd}. Despite this, directly measuring the absolute mass scale of neutrinos remains a formidable challenge in particle physics experiments. Neutrino mass has implications for the cosmic microwave background (CMB) and the universe's large-scale structure \cite{TopicalConvenersKNAbazajianJECarlstromATLee:2013bxd}, which makes cosmological observations a crucial approach for determining their absolute mass.

Current principal cosmological observations fall into two categories. Firstly, there are observations related to the universe's expansion history, such as baryon acoustic oscillations (BAO) and type Ia supernovae (SN). Secondly, observations concerning the cosmic structure's growth, such as redshift space distortions, weak gravitational lensing, and galaxy clusters counts, are also pivotal. These observations help constrain the total neutrino mass, as extensively discussed in Refs.~\cite{Hu:1997mj,Reid:2009nq,Li:2012vn,Li:2012spm,Wang:2012vh,Audren:2012vy,Riemer-Sorensen:2013jsa,Zhang:2014nta,Zhou:2014fva,Zhang:2014ifa,
Palanque-Delabrouille:2014jca,Zhang:2015rha,Geng:2015haa,Chen:2015oga,Zhang:2015uhk,Huang:2015wrx,Chen:2016eyp,Moresco:2016nqq,Lu:2016hsd,Wang:2016tsz,
Zhao:2016ecj,Kumar:2016zpg,Xu:2016ddc,Guo:2017hea,Zhang:2017rbg,Li:2017iur,Wang:2017htc,Zhao:2017jma,Guo:2018gyo,Feng:2019mym,Zhang:2019ipd,Feng:2019jqa,
Zhang:2020mox,Li:2020gtk,Giusarma:2016phn,Vagnozzi:2017ovm,Vagnozzi:2018jhn,Giusarma:2018jei,Tanseri:2022zfe,Yang:2017amu,Yang:2020ope,Yang:2020tax,
Amiri:2021kpp,Bazvand:2023fkx,Pang:2023joc,Geng:2014yoa,Liu:2020vgn}.

Recently, Ref.~\cite{Li:2020gtk} explored the cosmological constraints on total neutrino mass within the framework of vacuum energy interacting with cold dark matter (I$\Lambda$CDM), employing current observations. That study suggested that considering a direct interaction might alter the upper limit of the total neutrino mass, indicating that interactions between vacuum and cold dark matter could influence the cosmological measurement of neutrino mass. Nevertheless, current cosmological observations do not yet tightly constrain the total neutrino mass \cite{Planck:2015bpv,Aghanim:2018eyx}. However, gravitational-wave (GW) standard siren observations show promise as a powerful cosmological probe to aid in constraining the total neutrino mass.

The era of multi-messenger astronomy was inaugurated by the observation of the binary neutrino star (BNS) merger event GW170817 and its electromagnetic (EM) counterparts \cite{LIGOScientific:2017vwq,LIGOScientific:2017ync}. Analyzing the GW waveform from such events directly yields the absolute luminosity distance, known as a standard siren. Additionally, identifying the EM counterpart of a GW source allows for the measurement of its redshift. Establishing the relationship between cosmic distances and redshifts paves the way for probing the universe’s expansion history. Thus, GW standard sirens are expected to significantly influence the estimation of cosmological parameters.

In the coming decades, third-generation (3G) ground-based GW detectors, such as Cosmic Explorer (CE) \cite{LIGOScientific:2016wof} and Einstein Telescope (ET) \cite{Punturo:2010zz}, are set to revolutionize the measurement of GW signals. These detectors are anticipated to form a network that will significantly enhance the capability to detect GW events. Ref.~\cite{Jin:2022tdf} presented a preliminary discussion on using future GW standard siren observations to weigh neutrinos within interacting dark energy (IDE) models. This analysis is based on an assumption that 1000 standard sirens could be detected over a 10-year period by either the ET or CE. While this estimate aligns with projections in other studies \cite{Guo:2017hea,Zhang:2020mox,Li:2020gtk}, variations in the redshift distribution could impact cosmological parameter estimations.

This work aims to explore the potential of 3G standard siren observations to measure neutrino mass in IDE models. We will analyze the detection strategy employing a network comprising the ET and two CEs---one in the US with a 40-km arm length and another in Australia with a 20-km arm length, collectively referred to as ET2CE. Additionally, we consider observations of $\gamma$-ray bursts (GRBs) from a detector similar to THESEUS, enhancing our capability to ascertain cosmological parameters.

In our analysis of IDE models, we focus exclusively on the I$\Lambda$CDM models, characterized by a vacuum energy equation of state parameter $w=-1$. In these models, the energy conservation equations for vacuum energy and cold dark matter are described as follows:
\begin{equation}\label{conservation1}
\dot{\rho}_{\rm de} = Q,
\end{equation}
\begin{equation}\label{conservation2}
\dot{\rho}_{\rm c} = -3 H \rho_{\rm c}-Q,
\end{equation}
where $\rho_{\rm de}$ and $\rho_{\rm c}$ represent the densities of dark energy and cold dark matter, respectively, $H$ is the Hubble parameter, the dot denotes the derivative with respect to cosmic time $t$, and $Q$ is the rate of energy transfer. In the field of IDE, one common assumption is that $Q$ is proportional to the density of either dark matter or dark energy, i.e., $Q=\beta H \rho_{\rm c}$ or $Q=\beta H \rho_{\rm de}$~\cite{Li:2009zs,Li:2010ak,Fu:2011ab,Zhang:2012uu,Wang:2014oga,Cui:2015ueu,Feng:2016djj,Feng:2017usu,Feng:2018yew,Li:2018ydj}, where $\beta$ is a dimensionless coupling parameter and $H$ facilitates computational ease. Another assumption posits $Q$ as $Q=\beta H_0 \rho_{\rm c}$ or $Q=\beta H_0 \rho_{\rm de}$~\cite{Zhang:2013lea,Li:2017usw,Feng:2019mym,Feng:2019jqa,Wang:2021kxc,Zhao:2022bpd,Li:2023gtu}, with the Hubble constant $H_0$ used for dimensional consistency. In this work, we consider two forms of the energy transfer rate: $Q_1=\beta H \rho_{\rm c}$ and $Q_2=\beta H_0 \rho_{\rm c}$. According to Eqs.~(\ref{conservation1}) and (\ref{conservation2}), a positive $\beta$ implies that cold dark matter decays into vacuum energy, a negative $\beta$ suggests that vacuum energy decays into cold dark matter, and $\beta = 0$ indicates no interaction between them.

This paper is organized as follows. In Section~\ref{sec2}, we describe the methodology in our analysis. In Section~\ref{sec3}, we report the constraint results and make some relevant discussions. In Section~\ref{sec4}, we present a conclusion of this work.

\section{Methodology}
\label{sec2}

The parameter space vector of the I$\Lambda$CDM model is $\{\omega_b,~\omega_c,~\tau,~100\theta_{\rm MC},~\beta,~\ln (10^{10}A_s),~n_s\}$, where $\omega_b$ and $\omega_c$ are the present density of baryons and cold dark matter, respectively, $\tau$ is the Thomson scattering optical depth due to reionization, $\theta_{\rm MC}$ (multiplied by 100) is the radio between the comoving sound horizon and angular diameter distance at the decoupling epoch, $\beta$ is the coupling parameter in the I$\Lambda$CDM model, $A_s$ is the amplitude of primordial scalar perturbation power spectrum, and $n_s$ is its power-law spectral index.
When neutrino mass is considered in the I$\Lambda$CDM model, one extra free parameter $\sum m_\nu$ should be involved in the calculation.

Because we add neutrino mass into the I$\Lambda$CDM model, the model considered in this paper is called the I$\Lambda$CDM+$\sum m_\nu$ model.
For convenience, in this paper, we use I$\Lambda$CDM1+$\sum m_\nu$ and I$\Lambda$CDM2+$\sum m_\nu$ to denote the corresponding $Q_1=\beta H \rho_{\rm c}$ and $Q_2=\beta H_0 \rho_{\rm c}$ models that involve neutrino mass, respectively. Thus, there are eight independent parameters in total for these I$\Lambda$CDM+$\sum m_\nu$ (I$\Lambda$CDM1+$\sum m_\nu$ and I$\Lambda$CDM2+$\sum m_\nu$) models.

It should be mentioned that, when we consider the situation of vacuum energy interacting with cold dark matter, then the vacuum energy is not a pure background any more, and in this case, we must consider the vacuum energy perturbations. Note that, in the calculation of the dark energy perturbation evolution, a problem of perturbation instability appears; namely, the cosmological perturbations of dark energy will be divergent in some parts of the parameter space, which ruins the IDE cosmology at the perturbation level. Under such a circumstance, to avoid the perturbation instability problem in the I$\Lambda$CDM model, in this work, we treat the vacuum energy perturbations based on the extended parameterized post-Friedmann (PPF) approach~\cite{Li:2014eha,Li:2014cee,Li:2023fdk} (for the original version of the PPF method, see Refs.~\cite{Hu:2008zd,Fang:2008sn}). For more information about the calculation of cosmological perturbations in the IDE models and the PPF approach, see Refs.~\cite{Li:2015vla,Zhang:2017ize,Feng:2018yew,Guo:2017hea}.

In this paper, we simulate the GW standard siren data from the 3G GW detectors and future GRB detector and use them to constrain neutrino mass in the I$\Lambda$CDM models. We will investigate whether the GW standard sirens can improve the constraint on neutrino mass.

To show the constraining capability of the simulated GW standard siren data, we consider two data combinations for comparison: CMB+BAO+SN (abbreviated as CBS) and CBS+GW.
For the CMB data, we use the CMB likelihood, including the TT, TE, and EE spectra at $l\geq 30$, low-$l$ temperature commander likelihood, and low-$l$ SimAll EE likelihood, from the Planck 2018 data release~\cite{Aghanim:2018eyx}.
For the BAO data, we use the measurements from the 6dFGS ($z_{\rm eff}=0.106$)~\cite{Beutler:2011hx}, SDSS-MG ($z_{\rm eff}=0.15$)~\cite{Ross:2014qpa}, and BOSS-DR12 ($z_{\rm eff}=$ 0.38, 0.51, and 0.61)~\cite{Alam:2016hwk}.
For the SN data, we use the Pantheon sample, which is comprised of 1048 data points~\cite{Scolnic:2017caz}.

For the GW data, we use the simulation method described in Ref.~\cite{Han:2023exn}. Here, we provide only a brief overview. It is important to note that our approach includes a comprehensive calculation of the redshift distribution of GW-GRB events. This differs from the method used in previous studies, which assumed the detection of 1000 bright sirens over a 10-year observation period, as discussed in Refs.~\cite{Li:2019ajo,Jin:2022tdf,Cai:2016sby,Cai:2017aea,Wang:2018lun,Zhang:2018byx,Zhang:2019ylr,Zhang:2019ple,Zhang:2019loq,Jin:2020hmc,Jin:2021pcv,Wu:2022dgy,Jin:2023zhi,Zhang:2023gye}.

Based on the star formation rate~\cite{Vitale:2018yhm,Belgacem:2019tbw,Yang:2021qge}, the merger rate in the observer frame is
\begin{equation}
\mathcal{R}_{\rm obs}(z)=\frac{\mathcal{R}_{\rm m}(z)}{1+z} \frac{dV(z)}{dz},
\end{equation}
where $dV(z)/dz$ is the comoving volume element. $\mathcal{R}_{\rm m}$ is the BNS merger rate in the source frame that is given by
\begin{equation}
\mathcal{R}_{\rm m}(z)=\int_{t_{\rm min}}^{t_{\rm max}} \mathcal{R}_{\rm f}[t(z)-t_{\rm d}] P(t_{\rm d})dt_{\rm d},
\end{equation}
where $\mathcal{R}_{\rm f}$ is the cosmic star formation rate, for which we adopt the Madau-Dickinson model of Ref.~\cite{Madau:2014bja}, $t(z)$ is the universe's age at the time of merger, $t_{\rm d}$ is the time delay, and $t_{\rm min}$ and $t_{\rm max}$ are the minimum and maximum delay time for a massive binary to evolve merger~\cite{Vitale:2018yhm}, respectively. Here, $t_{\rm min}=20$ Myr and $t_{\rm max}$ is the Hubble time. The overall normalization is fixed by requiring that the value of $\mathcal{R}_{\rm m}(z=0)=920~\rm Gpc^{-3}~yr^{-1}$ agrees with the local rate estimated from the O1 LIGO and O2 LIGO/Virgo observation run~\cite{Eichhorn:2018phj}, which is also in accordance with the latest O3 observation run~\cite{KAGRA:2021duu}. $P(t_{\rm d})$ is the distribution of the time delay $t_{\rm d}$, and we follow and adopt the exponential form~\cite{Vitale:2018yhm}
\begin{equation}
	P(t_{\rm d})=\frac{1}{\tau}{\rm exp}(-t_{\rm d}/\tau),
\end{equation}
with an e-fold time of $\tau=100$ Myr for $t_{\rm d}>t_{\rm min}=20$ Myr.
We simulate a catalog of BNS merger for 10-year observation.
For each simulated GW source, the location $(\theta,\phi)$, polarization angle $\psi$, coalescence phase $\psi_{\rm c}$, and cosine of the inclination angle $\iota$ are drawn from uniform distributions.
The mass of neutron star (NS) is assumed to be a Gaussian distribution, and the center value of the NS mass is $1.33~M_{\odot}$, with a standard deviation of $0.09~M_{\odot}$, where $M_{\odot}$ is the solar mass.

Under the stationary phase approximation (SPA)~\cite{Zhang:2017srh}, the Fourier transform of the frequency-domain waveform for a detector network (with $N$ detectors) can be written as~\cite{Wen:2010cr,Zhao:2017cbb}
\begin{equation}
\tilde{\boldsymbol{h}}(f)=e^{-i\boldsymbol\Phi}\boldsymbol h(f),
\end{equation}
where $\boldsymbol\Phi$ is the $N\times N$ diagonal matrix with $\Phi_{ij}=2\pi f\delta_{ij}(\boldsymbol{n\cdot r}_k)$, $\boldsymbol n$ is the GW propagation direction, and $\boldsymbol r_k$ is the location of the $k$-th detector.  $\boldsymbol h$($f$) is
\begin{equation}
\boldsymbol h(f)=\Big[\frac{h_1(f)}{\sqrt{S_{\rm {n},1}(f)}}, \frac{h_2(f)}{\sqrt{S_{\rm {n},2}(f)}}, \ldots,\frac{h_N(f)}{\sqrt{S_{{\rm n},N}(f)}}\Big ]^{\rm T},
\end{equation}
where $S_{\rm {n},k}(f)$ is the one-sided noise power spectral density of the $k$-th detector, and $h_{k}(f)$ is the frequency domain GW waveform.

We consider the waveform in the inspiralling stage for the non-spinning BNS system in this study, and we adopt the restricted Post-Newtonian (PN) approximation and calculate the waveform to the 3.5 PN order~\cite{Cutler:1992tc,Sathyaprakash:2009xs}. The Fourier transform of the GW waveform of the $k$-th detector can be expressed as
\begin{align}
	h_k(f)=&\mathcal A_k f^{-7/6}{\rm exp}
	\{i[2\pi f t_{\rm c}-\pi/4-2\psi_c+2\Psi(f/2)]\nonumber\\ &-\varphi_{k,(2,0)})\},
\end{align}
where the detailed forms of $\Psi(f/2)$ and $\varphi_{k,(2,0)}$ can be found in Refs.~\cite{Cutler:1992tc,Zhao:2017cbb}. $\mathcal A_k$ is the Fourier amplitude, which can be written as
\begin{align}
	\mathcal A_k=&\frac{1}{d_{\rm L}}\sqrt{(F_{+,k}(1+\cos^{2}\iota))^{2}+(2F_{\times,k}\cos\iota)^{2}}\nonumber\\ &\times\sqrt{5\pi/96}\pi^{-7/6}\mathcal M^{5/6}_{\rm chirp},
\end{align}
where $d_{\rm L}$ is the luminosity distance of the GW source, $F_{+,k}$ and $F_{\times,k}$ are the antenna response functions of the $k$-th GW detector, $\mathcal M_{\rm chirp}=(1+z)\eta^{3/5}M$ is the observed chirp mass, $M=m_1+m_2$ is the total mass of binary system with component masses $m_1$ and $m_2$, and $\eta=m_1 m_2/M^2$ is the symmetric mass ratio. In this study, we adopt the GW waveform in the frequency domain, and then the time $t$ is replaced by $t_{f}=t_{\rm c}-(5 / 256) \mathcal{M}_{\rm chirp}^{-5 / 3}(\pi f)^{-8 / 3}$~\cite{Cutler:1992tc,Zhao:2017cbb}, where $t_{\rm c}$ is the coalescence time.

To obtain the detection of GW events, we need to calculate the signal-to-noise ratio (SNR) for each GW event.  The combined SNR for the detection network of $N$ detectors is given by
\begin{equation}
\rho=(\tilde{\boldsymbol h}|\tilde{\boldsymbol h})^{1/2}.
\end{equation}
The inner product is defined as
\begin{equation}
(\boldsymbol a|\boldsymbol b)=2\int_{f_{\rm lower}}^{f_{\rm upper}}\{\boldsymbol a(f)\boldsymbol b^*(f)+\boldsymbol a^*(f)\boldsymbol b(f)\}{\rm d}f,
\end{equation}
where $f_{\rm lower}$ is the lower cutoff frequency ($f_{\rm lower}=1$ Hz for ET and $f_{\rm lower}=5$ Hz for CE), $f_{\rm upper}=2/(6^{3/2}2\pi M_{\rm obs})$ is the frequency at the last stable orbit with $M_{\rm obs}=(m_1+m_2)(1+z)$, $\boldsymbol a$ and $\boldsymbol b$ are column matrices of the same dimension, and the star represents the conjugate transpose operator.
Here, we adopt the SNR threshold to be 12 in the simulation, and we assume a running period of 10 years and duty cycle of 100\% for the GW detector network.

For a GRB detected in coincidence with a GW signal, we require the peak flux to be higher than the flux limit of the satellite. Based on the fitting results of  GRB170817A~\cite{Howell:2018nhu}, we adopt the Gassian structured jet profile model
\begin{equation}
L_{\rm iso}(\theta_{\rm v})=L_{\rm on}\exp\left(-\frac{\theta^2_{\rm v}}{2\theta^2_{\rm c}} \right),
\label{eq:jet}
\end{equation}
where $L_{\rm iso}(\theta_{\rm v})$ is the isotropically equivalent luminosity of short GRB observed at different viewing angle $\theta_{\rm v}$, and we assume the jet's direction is aligned with the binary orbital angular momentum, namely, $\iota=\theta_{\rm v}$. $L_{\rm on}$ is the on-axis isotropic luminosity defined by $L_{\rm on}=L_{\rm iso}(0)$, and  $\theta_{\rm c}=4.7^{\circ}$ is the characteristic angle of the core.
The detection probability of a short GRB is determined by $\Phi(L){\rm d}L$. $\Phi(L)$ is the intrinsic luminosity function, given by

\begin{equation}
	\Phi(L)\propto
	\begin{cases}
		(L/L_*)^{\alpha}, & L<L_*, \\
		(L/L_*)^{\beta}, & L\ge L_*,
	\end{cases}
	\label{eq:distribution}
\end{equation}
where $L$ is the isotropic rest frame peak luminosity in the $1-10000$ keV energy range, $L_{*}$ is a characteristic value separating the two regimes, and $\alpha$ and $\beta$  are the characteristic slopes describing these regimes. In this paper, we use the values $L_{*}=2\times10^{52}$ erg sec$^{-1}$, $\alpha=-1.95$, and $\beta=-3$, which is in accordance with Ref.~\cite{Wanderman:2014eza}. In addition, we also assume a standard low-end cutoff in luminosity of $L_{\rm min} = 10^{49}$ erg sec$^{-1}$. Here, we term the on-axis isotropic luminosity $L_{\rm on}$ as the peak luminosity $L$.
According to the relation between luminosity and flux for GRB~\cite{Meszaros:1995dj,Meszaros:2011zr}, we can convert the flux limit $P_{\rm T}$ to the luminosity $L_{\rm iso}$~\cite{Han:2023exn}, where $P_{\rm T}=0.2~\rm ph~s^{-1}~cm^{-2}$ is the 50-300 keV band for THESEUS-like telescope~\cite{Stratta:2018ldl}. For the THESEUS-like telescope, we make the assumption of an 80\% duty cycle.
From the GW catalogue which has SNR larger than the threshold 12, then we can calculate the probability of the GRB detection for every GW event according to the probability distribution $\Phi(L){\rm d}L$.

For the total uncertainty of the luminosity distance $d_{\rm L}$, we consider the instrument error $\sigma_{d_{\rm L}}^{\rm inst}$, the weak-lensing error $\sigma_{d_{\rm L}}^{\rm lens}$, and the peculiar velocity error $\sigma_{d_{\rm L}}^{\rm pv}$, and thus the total error is given by
\begin{align}
\sigma_{d_{\rm L}}&~~=\sqrt{(\sigma_{d_{\rm L}}^{\rm inst})^2+(\sigma_{d_{\rm L}}^{\rm lens})^2+(\sigma_{d_{\rm L}}^{\rm pv})^2}.\label{total}
\end{align}

We first estimate the instrument error $\sigma_{d_{\rm L}}^{\rm inst}$ by using the Fisher information matrix. The Fisher information matrix of a network including $N$ independent GW detectors can be written as
\begin{equation}
F_{ij}=\left(\frac{\partial \tilde{\boldsymbol{h}}}{\partial \theta_i}\Bigg |\frac{\partial \tilde{\boldsymbol{h}}}{\partial \theta_j}\right)
\end{equation}
where $\theta_i$ denotes nine parameters ($d_{\rm L}$, $\mathcal{M}_{\rm chirp}$, $\eta$, $\theta$, $\phi$, $\iota$, $t_{\rm c}$, $\psi_{\rm c}$, $\psi$) for a GW event. The instrumental error of the parameter $\theta_i$ can be calculated by $\Delta \theta_i =\sqrt{(F^{-1})_{ii}}$, where $F_{ij}$ is the total Fisher information matrix for the network of $N$ interferometers.

The error caused by weak lensing is adopted from Refs.~\cite{Hirata:2010ba,Speri:2020hwc}
\begin{align}
\sigma_{d_{\rm L}}^{\rm lens}(z)=&\left[1-\frac{0.3}{\pi/2} \arctan(z/0.073)\right]\times d_{\rm L}(z)\nonumber\\ &\times 0.066\left [\frac{1-(1+z)^{-0.25}}{0.25}\right ]^{1.8}.\label{lens}
\end{align}

The error caused by the peculiar velocity of the GW source is given by~\cite{Kocsis:2005vv}
\begin{equation}
	\sigma_{d_{\rm L}}^{\rm pv}(z)=d_{\rm L}(z)\times \left [ 1+ \frac{c(1+z)^2}{H(z)d_{\rm L}(z)}\right ]\frac{\sqrt{\langle v^2\rangle}}{c},\label{pv}
\end{equation}
where $H(z)$ is the Hubble parameter, $c$ is the speed of light in a vacuum, and $\sqrt{\langle v^2\rangle}$ is the peculiar velocity of the GW source; we roughly use $\sqrt{\langle v^2\rangle}=500\ {\rm km\ s^{-1}}$.

In this work, we wish to study the constraint on neutrino mass in the I$\Lambda$CDM+$\sum m_\nu$ models from GW standard sirens.
As the first attempt to explore the impacts of GW-GRB joint observation on the neutrino mass, we do not consider all the different cases of 3G GW observations, i.e., single ET, single CE, the CE-CE network (one in the US and another in Australia), and the ET-CE-CE network (one ET detector and two CE-like detectors). Instead, we only consider one case of ET-CE-CE network (abbreviated as GW hereafter) as a concrete example to analyze.

In the following, we use these observational data to place constraints on these I$\Lambda$CDM+$\sum m_\nu$ models. We will use two data combinations, i.e., CBS and CBS+GW, to constrain the neutrino mass $\sum m_\nu$ and other cosmological parameters. We employ the modified version of the publicly available Markov-Chain Monte Corlo (MCMC) package CosmoMC~\cite{Lewis:2002ah} to perform the calculations, and we use the PPF package~\cite{Li:2014eha,Li:2014cee} for the I$\Lambda$CDM models to handle the perturbations of vacuum energy.

\section{Results and discussion}\label{sec3}

\begin{figure*}[!htp]
\includegraphics[scale=0.9]{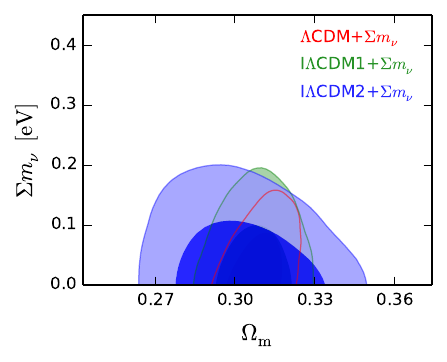}
\includegraphics[scale=0.9]{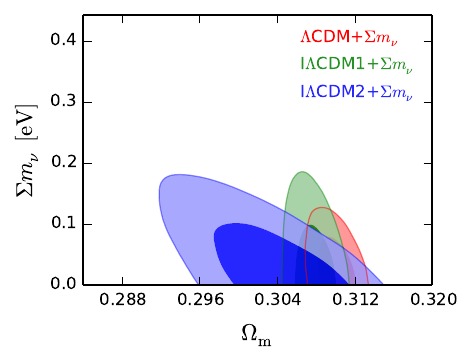}
\centering
 \caption{\label{fig1} (color online) Two-dimensional marginalized posterior contours (68.3\% and 95.4\% CL) in the $\Omega_{\rm m}$--$\sum m_\nu$ plane for the I$\Lambda$CDM+$\sum m_\nu$ models using the CBS (left) and CBS+GW (right) data combinations. }
\end{figure*}

\begin{figure*}[!htp]

\includegraphics[scale=0.5]{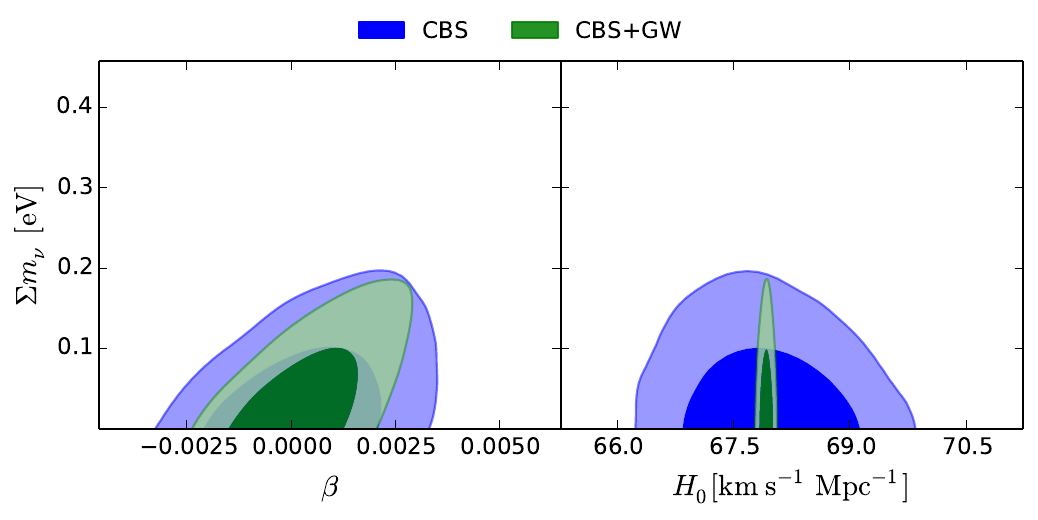}
\includegraphics[scale=0.5]{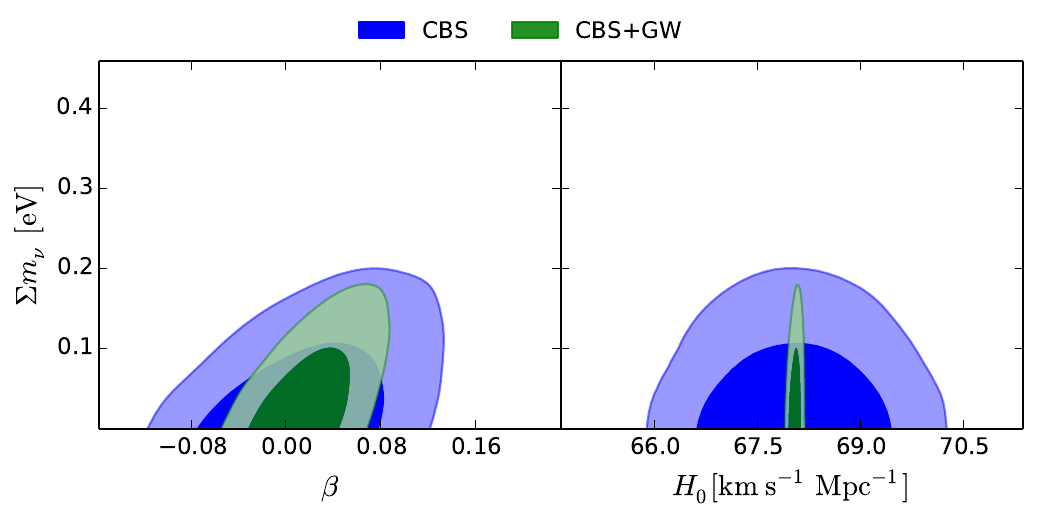}

\centering
 \caption{\label{fig2}(color online) Two-dimensional marginalized posterior contours (68.3\% and 95.4\% CL) in the $\sum m_\nu$--$\beta$ and $\sum m_\nu$--$H_0$ planes for the I$\Lambda$CDM1+$\sum m_\nu$($Q_1=\beta H\rho_{\rm c}$) (left) and I$\Lambda$CDM2+$\sum m_\nu$($Q_2=\beta H_0\rho_{\rm c}$) (right) models using the CBS and CBS+GW data combinations. }
\end{figure*}

In this section, we report the fitting results of these I$\Lambda$CDM+$\sum m_\nu$ models and discuss the impacts of GW standard siren observation on constraining neutrino mass.
We found that only approximately 640 standard sirens can be detected for the ET2CE+THESEUS network in 10 years, whose redshift distribution is given in Ref.~\cite{Han:2023exn}.
The fitting results are presented in Tables~\ref{tab1} and~\ref{tab2} as well as Figs.~\ref{fig1} and \ref{fig2}. In Table~\ref{tab1},  we quote $\pm 1\sigma$ errors for the parameters, but for the parameters that cannot be well-constrained, e.g., neutrino mass $\sum m_\nu$, we quote the 95.4\% CL upper limits.

\begin{table*}\small
\setlength\tabcolsep{3.3pt}
\renewcommand{\arraystretch}{1.5}
\caption{\label{tab1}Fitting results for the I$\Lambda$CDM+$\sum m_\nu$ models using the CBS and CBS+GW data. We quote $\pm 1\sigma$ errors, but for the parameters that cannot be well-constrained, we quote the 95.4\% CL upper limits. Here, $H_0$ is in units of ${\rm km}\ {\rm s^{-1}}\  {\rm Mpc^{-1}}$, and $\sum m_\nu$ is in units of eV.}
\centering
\begin{tabular}{ccccccccccccccccccc}

\hline \multicolumn{1}{c}{}&& \multicolumn{2}{c}{$\Lambda$CDM+$\sum m_\nu$} ($Q=0$) &&\multicolumn{2}{c}{I$\Lambda$CDM1+$\sum m_\nu$} ($Q_1=\beta H\rho_{\rm c}$)&&\multicolumn{2}{c}{I$\Lambda$CDM2+$\sum m_\nu$} ($Q_2=\beta H_0\rho_{\rm c}$)&\\
           \cline{3-4}\cline{6-7}\cline{9-10}
Parameter  && CBS & CBS+GW && CBS & CBS+GW && CBS & CBS+GW \\

\hline
$\Omega_m$&&$0.3097^{+0.0062}_{-0.0067}$&$0.3098\pm0.0012$&&$0.3078\pm0.0081$&$0.3076\pm0.0013$&&$0.3030^{+0.0160}_{-0.0180}$&$0.3024^{+0.0048}_{-0.0041}$\\
$\sigma_8$&&$0.8124^{+0.0128}_{-0.0088}$&$0.8136^{+0.0117}_{-0.0086}$&&$0.8140^{+0.0140}_{-0.0130}$&$0.8150^{+0.0130}_{-0.0110}$&&$0.8190\pm0.0200$&$0.8200\pm0.0140$\\
$H_0$&&$67.75^{+0.52}_{-0.48}$&$67.75\pm0.05$&&$67.92\pm0.65$&$67.92\pm0.05$&&$68.06^{+0.81}_{-0.82}$&$68.06\pm0.06$\\
$\beta$&&...&...&&$0.0005\pm0.0013$&$0.0005^{+0.0009}_{-0.0011}$&&$0.0230^{+0.0470}_{-0.0480}$&$0.0230^{+0.0260}_{-0.0290}$\\
$\sum m_\nu$&& $<0.123$&$<0.103$ && $<0.150$&$<0.143$ && $<0.156$&$<0.140$\\
\hline
\end{tabular}
\end{table*}

\begin{table*}\small
\setlength\tabcolsep{3.3pt}
\renewcommand{\arraystretch}{1.5}
\caption{\label{tab2}Absolute and relative errors of cosmological parameters in the $\Lambda$CDM+$\sum m_\nu$ and I$\Lambda$CDM+$\sum m_\nu$ models using the CBS and CBS+GW data combinations. Here, $H_0$ is in units of ${\rm km}\ {\rm s^{-1}}\  {\rm Mpc^{-1}}$. For a parameter $\xi$, $\sigma(\xi)$ and $\varepsilon(\xi)=\sigma(\xi)/\xi$ represent its absolute and relative errors, respectively.}
\centering
\begin{tabular}{ccccccccccccccccccc}

\hline \multicolumn{1}{c}{}&& \multicolumn{2}{c}{$\Lambda$CDM+$\sum m_\nu$} ($Q=0$) &&\multicolumn{2}{c}{I$\Lambda$CDM1+$\sum m_\nu$} ($Q_1=\beta H\rho_{\rm c}$)&&\multicolumn{2}{c}{I$\Lambda$CDM2+$\sum m_\nu$} ($Q_2=\beta H_0\rho_{\rm c}$)&\\
           \cline{3-4}\cline{6-7}\cline{9-10}
Parameter  && CBS & CBS+GW && CBS & CBS+GW && CBS & CBS+GW \\

\hline

$\sigma(\Omega_m)$&& $0.0065$&$0.0012$ && $0.0081$&$0.0013$ && $0.0170$&$0.0045$\\
$\sigma(\sigma_8)$&& $0.0108$&$0.0102$ && $0.0135$&$0.0120$ && $0.0200$&$0.0140$\\
$\sigma(H_0)$&& $0.50$&$0.05$ && $0.65$&$0.05$ && $0.82$&$0.06$\\
$\sigma(\beta)$&& $...$&$...$ && $0.0013$&$0.0010$ && $0.0475$&$0.0275$\\
$\varepsilon(\Omega_m)$&& $2.10\%$&$0.39\%$ && $2.63\%$&$0.42\%$ && $5.61\%$&$1.49\%$\\
$\varepsilon(\sigma_8)$&& $1.33\%$&$1.25\%$ && $1.66\%$&$1.47\%$ && $2.44\%$&$1.71\%$\\
$\varepsilon(H_0)$&& $0.74\%$&$0.07\%$ && $0.96\%$&$0.07\%$ && $1.20\%$&$0.09\%$\\

\hline
\end{tabular}
\end{table*}

First, we discuss the effect of simulated GW data of ET2CE on constraining the total neutrino mass. In the case of CBS constraints, from Table~\ref{tab1}, we have $\sum m_\nu<0.123$ eV for $\Lambda$CDM+$\sum m_\nu$, $\sum m_\nu<0.150$ eV for I$\Lambda$CDM1+$\sum m_\nu$, and $\sum m_\nu<0.156$ eV for I$\Lambda$CDM2+$\sum m_\nu$.
After adding the GW data of ET2CE, namely, using the data combination CBS+GW, the constraint results become $\sum m_\nu<0.103$ eV for $\Lambda$CDM+$\sum m_\nu$, $\sum m_\nu<0.143$ eV for I$\Lambda$CDM1+$\sum m_\nu$, and $\sum m_\nu<0.140$ eV for I$\Lambda$CDM2+$\sum m_\nu$.
We found that, compared with the $\Lambda$CDM+$\sum m_\nu$ model, the constraint on neutrino mass $\sum m_\nu$ becomes slighter looser in these I$\Lambda$CDM+$\sum m_\nu$ models.
We show the two-dimensional marginalized posterior contours (68.3\% and 95.4\% CL) in the $\Omega_{\rm m}$--$\sum m_\nu$ plane for the $\Lambda$CDM+$\sum m_\nu$ and I$\Lambda$CDM+$\sum m_\nu$ models in Fig.~\ref{fig1} (left panel for CBS and right panel for CBS+GW). From this figure, we can clearly see that once the interaction is considered in the model, the upper limit of $\sum m_\nu$ becomes larger.

Moreover, we also found that the GW data help reduce the upper limits of neutrino mass $\sum m_\nu$ by 16.26\%, 4.67\%, and 10.26\% for the $\Lambda$CDM+$\sum m_\nu$, I$\Lambda$CDM1+$\sum m_\nu$, and I$\Lambda$CDM2+$\sum m_\nu$ models, respectively. When the GW data are considered, the constraints on $\sum m_\nu$ become slightly tighter in the $\Lambda$CDM+$\sum m_\nu$ and I$\Lambda$CDM+$\sum m_\nu$ models (also see Fig.~\ref{fig2}).
Thus, in the I$\Lambda$CDM+$\sum m_\nu$ models, compared to the current observations, the GW standard siren observations from ET2CE can slightly improve the constraint on neutrino mass, which is in accordance with the conclusion on the IDE models in the previous study~\cite{Jin:2022tdf}.

Next, we discuss how the GW data help improve the constraints on other parameters, i.e., $\Omega_m$, $\sigma_8$, and $H_0$.
In Table~\ref{tab2}, we show the absolute and relative errors of $\Omega_m$, $\sigma_8$, and $H_0$ from the CBS and CBS+GW data combinations. Comparing the results from two data combinations, we found that the constraints on $\Omega_m$, $\sigma_8$, and $H_0$ are improved by 73.53\%--83.95\%, 5.56\%--30\%, and 90\%--92.68\%, respectively, when considering the ET2CE data. Obviously, the GW data can indeed effectively improve the constraints on the parameters $\Omega_m$ and $H_0$ (also see Figs.~\ref{fig1} and ~\ref{fig2}).

Finally, we present the constraint results of the coupling constant $\beta$.
By using CBS data, we have $\beta=0.0005\pm0.0013$ for the I$\Lambda$CDM1+$\sum m_\nu$ model and $\beta=0.0230^{+0.0470}_{-0.0480}$ for I$\Lambda$CDM2+$\sum m_\nu$ model. It is shown that a positive value of $\beta$ is slightly favored and $\beta>0$ is only at $0.38\sigma$ and $0.48\sigma$ levels, respectively.
By using CBS+GW data, we have $\beta=0.0005^{+0.0009}_{-0.0011}$ for the I$\Lambda$CDM1+$\sum m_\nu$ model and $\beta=0.0230^{+0.0260}_{-0.0290}$ for the I$\Lambda$CDM2+$\sum m_\nu$ model. Thus, we found that the error is slightly shrunk and now $\beta>0$ is favored at $0.45\sigma$ and $0.79\sigma$ levels, respectively.
For these two I$\Lambda$CDM+$\sum m_\nu$ models, one can clearly see that, in the I$\Lambda$CDM+$\sum m_\nu$ cases (with $Q_1=\beta H\rho_{\rm c}$ and $Q_2=\beta H_0\rho_{\rm c}$), $\beta=0$ is favored within the $1\sigma$ significance range, no matter whether the GW data are taken into account. Thus, in the two I$\Lambda$CDM+$\sum m_\nu$ models, there is no evidence of a nonzero interaction.
Moreover, comparing the values of $\beta$ from the two data combinations, we also found that the accuracy of $\beta$ is increased by 23.08\% ($Q_1=\beta H \rho_{\rm c}$) and 42.11\% ($Q_2=\beta H_0\rho_{\rm c}$) when adding the GW data of ET2CE in the cosmological fit (also see Fig.~\ref{fig2}).
This indicates that the GW data of the ET2CE can also improve the constraint accuracies of coupling parameter $\beta$.

\section{Conclusion}\label{sec4}

In this paper, we investigated how GW standard sirens from 3G ground-based GW detectors can constrain the total neutrino mass ($\sum m_\nu$) in interacting vacuum energy models with energy transfer forms $Q_1=\beta H\rho_{\rm c}$ and $Q_2=\beta H_0\rho_{\rm c}$. We used an extended version of the PPF approach to compute the vacuum energy perturbations within these interacting scenarios. We focused on the synergy between 3G GW detectors and a GRB detector for multi-messenger observations, specifically employing the ET2CE network as a representative for GW discussions. To evaluate the impact of GW data on the constraints of $\sum m_\nu$, we also considered existing CMB+BAO+SN data for comparison and combination.

Our analysis revealed that GW data can reduce the upper limits of $\sum m_\nu$ by 16.26\%, 4.67\%, and 10.26\% for the $\Lambda$CDM+$\sum m_\nu$, I$\Lambda$CDM1+$\sum m_\nu$, and I$\Lambda$CDM2+$\sum m_\nu$ models, respectively. Thus, GW standard siren data from ET2CE offer a modest improvement in constraining $\sum m_\nu$ compared to CBS alone. For the derived parameters $\Omega_m$ and $H_0$, incorporating GW data into the cosmological fit substantially enhances the precision: the accuracy of $\Omega_m$ improved by 73.53\%--83.95\% and that of $H_0$ by 90\%--92.68\%. These significant enhancements underscore the value of including GW data from ET2CE. Moreover, based on the combined constraints from CBS and CBS+GW data, we found that GW data from ET2CE also significantly refine the accuracy of the coupling strength $\beta$.

\begin{acknowledgments}
We thank Shang-Jie Jin for helpful discussions.
This work was supported by the National Natural Science Foundation of China (12305069, 11947022, 11975072, 11875102, 11835009),
the National SKA Program of China (2022SKA0110200, 2022SKA0110203), the National 111 Project (B16009),
and the Program of the Education Department of Liaoning Province (JYTMS20231695).
\end{acknowledgments}



\begin{thebibliography}{99}

\bibitem{Lesgourgues:2006nd}
J.~Lesgourgues and S.~Pastor,
Phys. Rept. \textbf{429} (2006), 307-379
[arXiv:astro-ph/0603494 [astro-ph]].

\bibitem{TopicalConvenersKNAbazajianJECarlstromATLee:2013bxd}
K.~N.~Abazajian \textit{et al.} [Topical Conveners: K.N. Abazajian, J.E. Carlstrom, A.T. Lee],
Astropart. Phys. \textbf{63} (2015), 66-80
[arXiv:1309.5383 [astro-ph.CO]].


\bibitem{Hu:1997mj}
W.~Hu, D.~J.~Eisenstein and M.~Tegmark,
Phys. Rev. Lett. \textbf{80} (1998), 5255-5258
[arXiv:astro-ph/9712057 [astro-ph]].

\bibitem{Reid:2009nq}
B.~A.~Reid, L.~Verde, R.~Jimenez and O.~Mena,
JCAP \textbf{01} (2010), 003
[arXiv:0910.0008 [astro-ph.CO]].

\bibitem{Li:2012vn}
H.~Li and X.~Zhang,
Phys. Lett. B \textbf{713} (2012), 160-164
[arXiv:1202.4071 [astro-ph.CO]].

\bibitem{Li:2012spm}
Y.~H.~Li, S.~Wang, X.~D.~Li and X.~Zhang,
JCAP \textbf{02} (2013), 033
[arXiv:1207.6679 [astro-ph.CO]].

\bibitem{Wang:2012vh}
X.~Wang, X.~L.~Meng, T.~J.~Zhang, H.~Shan, Y.~Gong, C.~Tao, X.~Chen and Y.~F.~Huang,
JCAP \textbf{11} (2012), 018
[arXiv:1210.2136 [astro-ph.CO]].

\bibitem{Audren:2012vy}
B.~Audren, J.~Lesgourgues, S.~Bird, M.~G.~Haehnelt and M.~Viel,
JCAP \textbf{01} (2013), 026
[arXiv:1210.2194 [astro-ph.CO]].

\bibitem{Riemer-Sorensen:2013jsa}
S.~Riemer-S\o{}rensen, D.~Parkinson and T.~M.~Davis,
Phys. Rev. D \textbf{89} (2014), 103505
[arXiv:1306.4153 [astro-ph.CO]].

\bibitem{Zhang:2014nta}
J.~F.~Zhang, Y.~H.~Li and X.~Zhang,
Eur. Phys. J. C \textbf{74} (2014), 2954
[arXiv:1404.3598 [astro-ph.CO]].

\bibitem{Zhou:2014fva}
X.~Y.~Zhou and J.~H.~He,
Commun. Theor. Phys. \textbf{62} (2014), 102-108
[arXiv:1406.6822 [astro-ph.CO]].

\bibitem{Zhang:2014ifa}
J.~F.~Zhang, J.~J.~Geng and X.~Zhang,
JCAP \textbf{10} (2014), 044
[arXiv:1408.0481 [astro-ph.CO]].

\bibitem{Palanque-Delabrouille:2014jca}
N.~Palanque-Delabrouille, C.~Y\`eche, J.~Lesgourgues, G.~Rossi, A.~Borde, M.~Viel, E.~Aubourg, D.~Kirkby, J.~M.~LeGoff and J.~Rich, \textit{et al.}
JCAP \textbf{02} (2015), 045
[arXiv:1410.7244 [astro-ph.CO]].

\bibitem{Geng:2014yoa}
C.~Q.~Geng, C.~C.~Lee and J.~L.~Shen,
Phys. Lett. B \textbf{740} (2015), 285-290
[arXiv:1411.3813 [astro-ph.CO]].

\bibitem{Zhang:2015rha}
J.~F.~Zhang, M.~M.~Zhao, Y.~H.~Li and X.~Zhang,
JCAP \textbf{04} (2015), 038
[arXiv:1502.04028 [astro-ph.CO]].

\bibitem{Geng:2015haa}
C.~Q.~Geng, C.~C.~Lee, R.~Myrzakulov, M.~Sami and E.~N.~Saridakis,
JCAP \textbf{01} (2016), 049
[arXiv:1504.08141 [astro-ph.CO]].

\bibitem{Chen:2015oga}
Y.~Chen and L.~Xu,
Phys. Lett. B \textbf{752} (2016), 66-75
[arXiv:1507.02008 [astro-ph.CO]].

\bibitem{Zhang:2015uhk}
X.~Zhang,
Phys. Rev. D \textbf{93} (2016) no.8, 083011
[arXiv:1511.02651 [astro-ph.CO]].

\bibitem{Huang:2015wrx}
Q.~G.~Huang, K.~Wang and S.~Wang,
Eur. Phys. J. C \textbf{76} (2016) no.9, 489
[arXiv:1512.05899 [astro-ph.CO]].

\bibitem{Chen:2016eyp}
Y.~Chen, B.~Ratra, M.~Biesiada, S.~Li and Z.~H.~Zhu,
Astrophys. J. \textbf{829} (2016) no.2, 61
[arXiv:1603.07115 [astro-ph.CO]].

\bibitem{Moresco:2016nqq}
M.~Moresco, R.~Jimenez, L.~Verde, A.~Cimatti, L.~Pozzetti, C.~Maraston and D.~Thomas,
JCAP \textbf{12} (2016), 039
[arXiv:1604.00183 [astro-ph.CO]].

\bibitem{Giusarma:2016phn}
E.~Giusarma, M.~Gerbino, O.~Mena, S.~Vagnozzi, S.~Ho and K.~Freese,
Phys. Rev. D \textbf{94} (2016) no.8, 083522
[arXiv:1605.04320 [astro-ph.CO]].

\bibitem{Lu:2016hsd}
J.~Lu, M.~Liu, Y.~Wu, Y.~Wang and W.~Yang,
Eur. Phys. J. C \textbf{76} (2016) no.12, 679
[arXiv:1606.02987 [astro-ph.CO]].

\bibitem{Wang:2016tsz}
S.~Wang, Y.~F.~Wang, D.~M.~Xia and X.~Zhang,
Phys. Rev. D \textbf{94} (2016) no.8, 083519
[arXiv:1608.00672 [astro-ph.CO]].

\bibitem{Zhao:2016ecj}
M.~M.~Zhao, Y.~H.~Li, J.~F.~Zhang and X.~Zhang,
Mon. Not. Roy. Astron. Soc. \textbf{469} (2017) no.2, 1713-1724
[arXiv:1608.01219 [astro-ph.CO]].

\bibitem{Kumar:2016zpg}
S.~Kumar and R.~C.~Nunes,
Phys. Rev. D \textbf{94} (2016) no.12, 123511
[arXiv:1608.02454 [astro-ph.CO]].

\bibitem{Xu:2016ddc}
L.~Xu and Q.~G.~Huang,
Sci. China Phys. Mech. Astron. \textbf{61} (2018) no.3, 039521
[arXiv:1611.05178 [astro-ph.CO]].

\bibitem{Vagnozzi:2017ovm}
S.~Vagnozzi, E.~Giusarma, O.~Mena, K.~Freese, M.~Gerbino, S.~Ho and M.~Lattanzi,
Phys. Rev. D \textbf{96} (2017) no.12, 123503
[arXiv:1701.08172 [astro-ph.CO]].

\bibitem{Guo:2017hea}
R.~Y.~Guo, Y.~H.~Li, J.~F.~Zhang and X.~Zhang,
JCAP \textbf{05} (2017), 040
[arXiv:1702.04189 [astro-ph.CO]].

\bibitem{Zhang:2017rbg}
X.~Zhang,
Sci. China Phys. Mech. Astron. \textbf{60} (2017) no.6, 060431
[arXiv:1703.00651 [astro-ph.CO]].

\bibitem{Li:2017iur}
E.~K.~Li, H.~Zhang, M.~Du, Z.~H.~Zhou and L.~Xu,
JCAP \textbf{08} (2018), 042
[arXiv:1703.01554 [astro-ph.CO]].

\bibitem{Yang:2017amu}
W.~Yang, R.~C.~Nunes, S.~Pan and D.~F.~Mota,
Phys. Rev. D \textbf{95} (2017) no.10, 103522
[arXiv:1703.02556 [astro-ph.CO]].

\bibitem{Wang:2017htc}
S.~Wang, Y.~F.~Wang and D.~M.~Xia,
Chin. Phys. C \textbf{42} (2018) no.6, 065103
[arXiv:1707.00588 [astro-ph.CO]].

\bibitem{Zhao:2017jma}
M.~M.~Zhao, J.~F.~Zhang and X.~Zhang,
Phys. Lett. B \textbf{779} (2018), 473-478
[arXiv:1710.02391 [astro-ph.CO]].

\bibitem{Vagnozzi:2018jhn}
S.~Vagnozzi, S.~Dhawan, M.~Gerbino, K.~Freese, A.~Goobar and O.~Mena,
Phys. Rev. D \textbf{98} (2018) no.8, 083501
[arXiv:1801.08553 [astro-ph.CO]].

\bibitem{Giusarma:2018jei}
E.~Giusarma, S.~Vagnozzi, S.~Ho, S.~Ferraro, K.~Freese, R.~Kamen-Rubio and K.~B.~Luk,
Phys. Rev. D \textbf{98} (2018) no.12, 123526
[arXiv:1802.08694 [astro-ph.CO]].

\bibitem{Guo:2018gyo}
R.~Y.~Guo, J.~F.~Zhang and X.~Zhang,
Chin. Phys. C \textbf{42} (2018) no.9, 095103
[arXiv:1803.06910 [astro-ph.CO]].

\bibitem{Feng:2019mym}
L.~Feng, H.~L.~Li, J.~F.~Zhang and X.~Zhang,
Sci. China Phys. Mech. Astron. \textbf{63} (2020) no.2, 220401
[arXiv:1903.08848 [astro-ph.CO]].

\bibitem{Zhang:2019ipd}
J.~F.~Zhang, B.~Wang and X.~Zhang,
Sci. China Phys. Mech. Astron. \textbf{63} (2020) no.8, 280411
[arXiv:1907.00179 [astro-ph.CO]].

\bibitem{Feng:2019jqa}
L.~Feng, D.~Z.~He, H.~L.~Li, J.~F.~Zhang and X.~Zhang,
Sci. China Phys. Mech. Astron. \textbf{63} (2020) no.9, 290404
[arXiv:1910.03872 [astro-ph.CO]].

\bibitem{Liu:2020vgn}
Z.~Liu and H.~Miao,
Int. J. Mod. Phys. D \textbf{29} (2020) no.13, 2050088
[arXiv:2002.05563 [astro-ph.CO]].

\bibitem{Yang:2020tax}
W.~Yang, E.~Di Valentino, O.~Mena and S.~Pan,
Phys. Rev. D \textbf{102} (2020) no.2, 023535
[arXiv:2003.12552 [astro-ph.CO]].

\bibitem{Zhang:2020mox}
M.~Zhang, J.~F.~Zhang and X.~Zhang,
Commun. Theor. Phys. \textbf{72} (2020) no.12, 125402
[arXiv:2005.04647 [astro-ph.CO]].

\bibitem{Li:2020gtk}
H.~L.~Li, J.~F.~Zhang and X.~Zhang,
Commun. Theor. Phys. \textbf{72} (2020) no.12, 125401
[arXiv:2005.12041 [astro-ph.CO]].

\bibitem{Yang:2020ope}
W.~Yang, E.~Di Valentino, S.~Pan and O.~Mena,
Phys. Dark Univ. \textbf{31} (2021), 100762
[arXiv:2007.02927 [astro-ph.CO]].

\bibitem{Amiri:2021kpp}
H.~R.~Amiri, A.~Salehi and A.~H.~Noroozi,
Eur. Phys. J. C \textbf{81} (2021) no.5, 479

\bibitem{Tanseri:2022zfe}
I.~Tanseri, S.~Hagstotz, S.~Vagnozzi, E.~Giusarma and K.~Freese,
JHEAp \textbf{36} (2022), 1-26
[arXiv:2207.01913 [astro-ph.CO]].

\bibitem{Bazvand:2023fkx}
P.~Bazvand, A.~Salehi and R.~Sepahvand,
Eur. Phys. J. Plus \textbf{138} (2023) no.5, 448

\bibitem{Pang:2023joc}
Y.~H.~Pang, X.~Zhang and Q.~G.~Huang,
Chin. Phys. C \textbf{48} (2024) no.6, 065102
[arXiv:2312.07188 [astro-ph.CO]].

\bibitem{Planck:2015bpv}
N.~Aghanim \textit{et al.} [Planck],
Astron. Astrophys. \textbf{594} (2016), A11
[arXiv:1507.02704 [astro-ph.CO]].

\bibitem{Aghanim:2018eyx}
N.~Aghanim \textit{et al.} [Planck],
Astron. Astrophys. \textbf{641} (2020), A6
[arXiv:1807.06209 [astro-ph.CO]].

\bibitem{LIGOScientific:2017vwq}
B.~P.~Abbott \textit{et al.} [LIGO Scientific and Virgo],
Phys. Rev. Lett. \textbf{119} (2017) no.16, 161101
[arXiv:1710.05832 [gr-qc]].

\bibitem{LIGOScientific:2017ync}
B.~P.~Abbott \textit{et al.} [LIGO Scientific, Virgo, Fermi GBM, INTEGRAL, IceCube, AstroSat Cadmium Zinc Telluride Imager Team, IPN, Insight-Hxmt, ANTARES, Swift, AGILE Team, 1M2H Team, Dark Energy Camera GW-EM, DES, DLT40, GRAWITA, Fermi-LAT, ATCA, ASKAP, Las Cumbres Observatory Group, OzGrav, DWF (Deeper Wider Faster Program), AST3, CAASTRO, VINROUGE, MASTER, J-GEM, GROWTH, JAGWAR, CaltechNRAO, TTU-NRAO, NuSTAR, Pan-STARRS, MAXI Team, TZAC Consortium, KU, Nordic Optical Telescope, ePESSTO, GROND, Texas Tech University, SALT Group, TOROS, BOOTES, MWA, CALET, IKI-GW Follow-up, H.E.S.S., LOFAR, LWA, HAWC, Pierre Auger, ALMA, Euro VLBI Team, Pi of Sky, Chandra Team at McGill University, DFN, ATLAS Telescopes, High Time Resolution Universe Survey, RIMAS, RATIR and SKA South Africa/MeerKAT],
Astrophys. J. Lett. \textbf{848} (2017) no.2, L12
[arXiv:1710.05833 [astro-ph.HE]].

\bibitem{LIGOScientific:2016wof}
B.~P.~Abbott \textit{et al.} [LIGO Scientific],
Class. Quant. Grav. \textbf{34} (2017) no.4, 044001
[arXiv:1607.08697 [astro-ph.IM]].

\bibitem{Punturo:2010zz}
M.~Punturo, M.~Abernathy, F.~Acernese, B.~Allen, N.~Andersson, K.~Arun, F.~Barone, B.~Barr, M.~Barsuglia and M.~Beker, \textit{et al.}
Class. Quant. Grav. \textbf{27} (2010), 194002
doi:10.1088/0264-9381/27/19/194002
\bibitem{Jin:2022tdf}
S.~J.~Jin, R.~Q.~Zhu, L.~F.~Wang, H.~L.~Li, J.~F.~Zhang and X.~Zhang,
Commun. Theor. Phys. \textbf{74} (2022) no.10, 105404
[arXiv:2204.04689 [astro-ph.CO]].


\bibitem{Li:2009zs}
M.~Li, X.~D.~Li, S.~Wang, Y.~Wang and X.~Zhang,
JCAP \textbf{12} (2009), 014
[arXiv:0910.3855 [astro-ph.CO]].

\bibitem{Li:2010ak}
Y.~Li, J.~Ma, J.~Cui, Z.~Wang and X.~Zhang,
Sci. China Phys. Mech. Astron. \textbf{54} (2011), 1367-1377
[arXiv:1011.6122 [astro-ph.CO]].


\bibitem{Fu:2011ab}
T.~F.~Fu, J.~F.~Zhang, J.~Q.~Chen and X.~Zhang,
Eur. Phys. J. C \textbf{72} (2012), 1932
[arXiv:1112.2350 [astro-ph.CO]].


\bibitem{Zhang:2012uu}
Z.~Zhang, S.~Li, X.~D.~Li, X.~Zhang and M.~Li,
JCAP \textbf{06} (2012), 009
[arXiv:1204.6135 [astro-ph.CO]].

\bibitem{Wang:2014oga}
S.~Wang, Y.~Z.~Wang, J.~J.~Geng and X.~Zhang,
Eur. Phys. J. C \textbf{74} (2014) no.11, 3148
[arXiv:1406.0072 [astro-ph.CO]].

\bibitem{Cui:2015ueu}
J.~L.~Cui, L.~Yin, L.~F.~Wang, Y.~H.~Li and X.~Zhang,
JCAP \textbf{09} (2015), 024
[arXiv:1503.08948 [astro-ph.CO]].

\bibitem{Feng:2016djj}
L.~Feng and X.~Zhang,
JCAP \textbf{08} (2016), 072
[arXiv:1607.05567 [astro-ph.CO]].

\bibitem{Feng:2017usu}
L.~Feng, J.~F.~Zhang and X.~Zhang,
Phys. Dark Univ. \textbf{23} (2019), 100261
[arXiv:1712.03148 [astro-ph.CO]].

\bibitem{Feng:2018yew}
L.~Feng, Y.~H.~Li, F.~Yu, J.~F.~Zhang and X.~Zhang,
Eur. Phys. J. C \textbf{78} (2018) no.10, 865
[arXiv:1807.03022 [astro-ph.CO]].

\bibitem{Li:2018ydj}
H.~L.~Li, L.~Feng, J.~F.~Zhang and X.~Zhang,
Sci. China Phys. Mech. Astron. \textbf{62} (2019) no.12, 120411
[arXiv:1812.00319 [astro-ph.CO]].

\bibitem{Zhang:2013lea}
J.~Zhang, L.~Zhao and X.~Zhang,
Sci. China Phys. Mech. Astron. \textbf{57} (2014), 387-392
[arXiv:1306.1289 [astro-ph.CO]].

\bibitem{Li:2017usw}
H.~L.~Li, J.~F.~Zhang, L.~Feng and X.~Zhang,
Eur. Phys. J. C \textbf{77} (2017) no.12, 907
[arXiv:1711.06159 [astro-ph.CO]].

\bibitem{Wang:2021kxc}
L.~F.~Wang, J.~H.~Zhang, D.~Z.~He, J.~F.~Zhang and X.~Zhang,
Mon. Not. Roy. Astron. Soc. \textbf{514} (2022) no.1, 1433-1440
[arXiv:2102.09331 [astro-ph.CO]].

\bibitem{Zhao:2022bpd}
Z.~W.~Zhao, L.~F.~Wang, J.~G.~Zhang, J.~F.~Zhang and X.~Zhang,
JCAP \textbf{04} (2023), 022
[arXiv:2210.07162 [astro-ph.CO]].

\bibitem{Li:2023gtu}
T.~N.~Li, S.~J.~Jin, H.~L.~Li, J.~F.~Zhang and X.~Zhang,
Astrophys. J. \textbf{963} (2024) no.1, 52
[arXiv:2310.15879 [astro-ph.CO]].

\bibitem{Li:2014eha}
Y.~H.~Li, J.~F.~Zhang and X.~Zhang,
Phys. Rev. D \textbf{90} (2014) no.6, 063005
[arXiv:1404.5220 [astro-ph.CO]].
\bibitem{Li:2014cee}
Y.~H.~Li, J.~F.~Zhang and X.~Zhang,
Phys. Rev. D \textbf{90} (2014) no.12, 123007
[arXiv:1409.7205 [astro-ph.CO]].

\bibitem{Li:2023fdk}
Y.~H.~Li and X.~Zhang,
JCAP \textbf{09} (2023), 046
[arXiv:2306.01593 [astro-ph.CO]].


\bibitem{Hu:2008zd}
W.~Hu,
Phys. Rev. D \textbf{77} (2008), 103524
[arXiv:0801.2433 [astro-ph]].

\bibitem{Fang:2008sn}
W.~Fang, W.~Hu and A.~Lewis,
Phys. Rev. D \textbf{78} (2008), 087303
[arXiv:0808.3125 [astro-ph]].

\bibitem{Li:2015vla}
Y.~H.~Li, J.~F.~Zhang and X.~Zhang,
Phys. Rev. D \textbf{93} (2016) no.2, 023002
[arXiv:1506.06349 [astro-ph.CO]].

\bibitem{Zhang:2017ize}
X.~Zhang,
Sci. China Phys. Mech. Astron. \textbf{60} (2017) no.5, 050431
[arXiv:1702.04564 [astro-ph.CO]].


\bibitem{Beutler:2011hx}
F.~Beutler, C.~Blake, M.~Colless, D.~H.~Jones, L.~Staveley-Smith, L.~Campbell, Q.~Parker, W.~Saunders and F.~Watson,
Mon. Not. Roy. Astron. Soc. \textbf{416} (2011), 3017-3032
[arXiv:1106.3366 [astro-ph.CO]].

\bibitem{Ross:2014qpa}
A.~J.~Ross, L.~Samushia, C.~Howlett, W.~J.~Percival, A.~Burden and M.~Manera,
Mon. Not. Roy. Astron. Soc. \textbf{449} (2015) no.1, 835-847
[arXiv:1409.3242 [astro-ph.CO]].

\bibitem{Alam:2016hwk}
S.~Alam \textit{et al.} [BOSS],
Mon. Not. Roy. Astron. Soc. \textbf{470} (2017) no.3, 2617-2652
[arXiv:1607.03155 [astro-ph.CO]].

\bibitem{Scolnic:2017caz}
D.~M.~Scolnic, D.~O.~Jones, A.~Rest, Y.~C.~Pan, R.~Chornock, R.~J.~Foley, M.~E.~Huber, R.~Kessler, G.~Narayan and A.~G.~Riess, \textit{et al.}
Astrophys. J. \textbf{859} (2018) no.2, 101
[arXiv:1710.00845 [astro-ph.CO]].

\bibitem{Han:2023exn}
T.~Han, S.~J.~Jin, J.~F.~Zhang and X.~Zhang,
Eur. Phys. J. C \textbf{84}, no.7, 663 (2024)
[arXiv:2309.14965 [astro-ph.CO]].


\bibitem{Cai:2016sby}
R.~G.~Cai and T.~Yang,
Phys. Rev. D \textbf{95} (2017) no.4, 044024
[arXiv:1608.08008 [astro-ph.CO]].

\bibitem{Cai:2017aea}
R.~G.~Cai, T.~B.~Liu, X.~W.~Liu, S.~J.~Wang and T.~Yang,
Phys. Rev. D \textbf{97} (2018) no.10, 103005
[arXiv:1712.00952 [astro-ph.CO]].

\bibitem{Wang:2018lun}
L.~F.~Wang, X.~N.~Zhang, J.~F.~Zhang and X.~Zhang,
Phys. Lett. B \textbf{782} (2018), 87-93
[arXiv:1802.04720 [astro-ph.CO]].

\bibitem{Zhang:2018byx}
X.~N.~Zhang, L.~F.~Wang, J.~F.~Zhang and X.~Zhang,
Phys. Rev. D \textbf{99} (2019) no.6, 063510
[arXiv:1804.08379 [astro-ph.CO]].

\bibitem{Zhang:2019ylr}
X.~Zhang,
Sci. China Phys. Mech. Astron. \textbf{62} (2019) no.11, 110431
[arXiv:1905.11122 [astro-ph.CO]].

\bibitem{Zhang:2019ple}
J.~F.~Zhang, H.~Y.~Dong, J.~Z.~Qi and X.~Zhang,
Eur. Phys. J. C \textbf{80} (2020) no.3, 217
[arXiv:1906.07504 [astro-ph.CO]].

\bibitem{Zhang:2019loq}
J.~F.~Zhang, M.~Zhang, S.~J.~Jin, J.~Z.~Qi and X.~Zhang,
JCAP \textbf{09} (2019), 068
[arXiv:1907.03238 [astro-ph.CO]].

\bibitem{Li:2019ajo}
H.~L.~Li, D.~Z.~He, J.~F.~Zhang and X.~Zhang,
JCAP \textbf{06} (2020), 038
[arXiv:1908.03098 [astro-ph.CO]].

\bibitem{Jin:2020hmc}
S.~J.~Jin, D.~Z.~He, Y.~Xu, J.~F.~Zhang and X.~Zhang,
JCAP \textbf{03} (2020), 051
[arXiv:2001.05393 [astro-ph.CO]].

\bibitem{Jin:2021pcv}
S.~J.~Jin, L.~F.~Wang, P.~J.~Wu, J.~F.~Zhang and X.~Zhang,
Phys. Rev. D \textbf{104} (2021) no.10, 103507
[arXiv:2106.01859 [astro-ph.CO]].

\bibitem{Wu:2022dgy}
P.~J.~Wu, Y.~Shao, S.~J.~Jin and X.~Zhang,
JCAP \textbf{06} (2023), 052
[arXiv:2202.09726 [astro-ph.CO]].

\bibitem{Jin:2023zhi}
S.~J.~Jin, S.~S.~Xing, Y.~Shao, J.~F.~Zhang and X.~Zhang,
Chin. Phys. C \textbf{47} (2023) no.6, 065104
[arXiv:2301.06722 [astro-ph.CO]].

\bibitem{Zhang:2023gye}
J.~G.~Zhang, Z.~W.~Zhao, Y.~Li, J.~F.~Zhang, D.~Li and X.~Zhang,
Sci. China Phys. Mech. Astron. \textbf{66} (2023) no.12, 120412
[arXiv:2307.01605 [astro-ph.CO]].

\bibitem{Vitale:2018yhm}
S.~Vitale, W.~M.~Farr, K.~Ng and C.~L.~Rodriguez,
Astrophys. J. Lett. \textbf{886} (2019) no.1, L1
[arXiv:1808.00901 [astro-ph.HE]].

\bibitem{Belgacem:2019tbw}
E.~Belgacem, Y.~Dirian, S.~Foffa, E.~J.~Howell, M.~Maggiore and T.~Regimbau,
JCAP \textbf{08} (2019), 015
[arXiv:1907.01487 [astro-ph.CO]].

\bibitem{Yang:2021qge}
T.~Yang,
JCAP \textbf{05} (2021), 044
[arXiv:2103.01923 [astro-ph.CO]].

\bibitem{Madau:2014bja}
P.~Madau and M.~Dickinson,
Ann. Rev. Astron. Astrophys. \textbf{52} (2014), 415-486
[arXiv:1403.0007 [astro-ph.CO]].

\bibitem{Eichhorn:2018phj}
A.~Eichhorn, T.~Koslowski and A.~D.~Pereira,
Universe \textbf{5} (2019) no.2, 53
[arXiv:1811.12909 [gr-qc]].

\bibitem{KAGRA:2021duu}
R.~Abbott \textit{et al.} [KAGRA, VIRGO and LIGO Scientific],
Phys. Rev. X \textbf{13} (2023) no.1, 011048
[arXiv:2111.03634 [astro-ph.HE]].


\bibitem{Zhang:2017srh}
X.~Zhang, T.~Liu and W.~Zhao,
Phys. Rev. D \textbf{95} (2017) no.10, 104027
[arXiv:1702.08752 [gr-qc]].

\bibitem{Wen:2010cr}
L.~Wen and Y.~Chen,
Phys. Rev. D \textbf{81} (2010), 082001
[arXiv:1003.2504 [astro-ph.CO]].

\bibitem{Zhao:2017cbb}
W.~Zhao and L.~Wen,
Phys. Rev. D \textbf{97} (2018) no.6, 064031
[arXiv:1710.05325 [astro-ph.CO]].

\bibitem{Cutler:1992tc}
C.~Cutler, T.~A.~Apostolatos, L.~Bildsten, L.~S.~Finn, E.~E.~Flanagan, D.~Kennefick, D.~M.~Markovic, A.~Ori, E.~Poisson and G.~J.~Sussman, \textit{et al.}
Phys. Rev. Lett. \textbf{70} (1993), 2984-2987
[arXiv:astro-ph/9208005 [astro-ph]].


\bibitem{Sathyaprakash:2009xs}
B.~S.~Sathyaprakash and B.~F.~Schutz,
Living Rev. Rel. \textbf{12} (2009), 2
[arXiv:0903.0338 [gr-qc]].

\bibitem{Howell:2018nhu}
E.~J.~Howell, K.~Ackley, A.~Rowlinson and D.~Coward,
[arXiv:1811.09168 [astro-ph.HE]].

\bibitem{Wanderman:2014eza}
D.~Wanderman and T.~Piran,
Mon. Not. Roy. Astron. Soc. \textbf{448} (2015) no.4, 3026-3037
[arXiv:1405.5878 [astro-ph.HE]].

\bibitem{Meszaros:1995dj}
P.~Meszaros and A.~Meszaros,
Astrophys. J. \textbf{449} (1995), 9-16
[arXiv:astro-ph/9503087 [astro-ph]].

\bibitem{Meszaros:2011zr}
A.~Meszaros, J.~Ripa and F.~Ryde,
Astron. Astrophys. \textbf{529} (2011), A55
[arXiv:1101.5040 [astro-ph.CO]].

\bibitem{Stratta:2018ldl}
G.~Stratta, L.~Amati, R.~Ciolfi and S.~Vinciguerra,
Mem. Soc. Ast. It. \textbf{89} (2018) no.2, 205-212
[arXiv:1802.01677 [astro-ph.IM]].


\bibitem{Hirata:2010ba}
C.~M.~Hirata, D.~E.~Holz and C.~Cutler,
Phys. Rev. D \textbf{81} (2010), 124046
[arXiv:1004.3988 [astro-ph.CO]].

\bibitem{Speri:2020hwc}
L.~Speri, N.~Tamanini, R.~R.~Caldwell, J.~R.~Gair and B.~Wang,
Phys. Rev. D \textbf{103} (2021) no.8, 083526
[arXiv:2010.09049 [astro-ph.CO]].

\bibitem{Kocsis:2005vv}
B.~Kocsis, Z.~Frei, Z.~Haiman and K.~Menou,
Astrophys. J. \textbf{637} (2006), 27-37
[arXiv:astro-ph/0505394 [astro-ph]].

\bibitem{Lewis:2002ah}
A.~Lewis and S.~Bridle,
Phys. Rev. D \textbf{66} (2002), 103511
[arXiv:astro-ph/0205436 [astro-ph]].
\end{thebibliography}
\end{document}